\documentclass[preprint,authoryear,5p,times,twocolumn]{elsarticle}

\usepackage{lineno}
\usepackage{amsmath,amssymb}
\usepackage[colorlinks=True,
            urlcolor=blue,
            linkcolor=blue,
            citecolor= blue,
            dvips]{hyperref}
\usepackage{journals} 
\usepackage{booktabs}

\journal{Icarus}

\begin{document}

\title{Scaling laws in spherical shell dynamos with free-slip boundaries}

\author{Rakesh K. Yadav}
\ead{yadav@mps.mpg.de}
\author{Thomas Gastine}
\author{Ulrich R. Christensen}

\address{Max-Planck-Institute f\"ur Sonensystemforchung, 37191, Katlenburg-Lindau, Germany}

\begin{abstract}
Numerical simulations of convection driven rotating spherical shell dynamos have often been performed with rigid boundary conditions, as is appropriate for the metallic cores of terrestrial planets. Free-slip boundaries are more appropriate for dynamos in other astrophysical objects, such as gas-giants or stars. Using a set of 57 direct numerical simulations, we investigate the effect of free-slip boundary conditions on the scaling properties of heat flow, flow velocity and magnetic field strength and compare it with earlier results for rigid boundaries.  We find that the nature of the mechanical boundary condition has only a minor influence on the scaling laws. We also find that although dipolar and multipolar dynamos exhibit approximately the same scaling exponents,  there is an offset in the scaling pre-factors for velocity and magnetic field strength. We argue that the offset can be attributed to the differences in the zonal flow contribution between dipolar and multipolar dynamos. 
\end{abstract}

\maketitle

\section{Introduction}

Numerical simulations of dynamos in geometries appropriate for the cores of terrestrial planets have greatly enhanced our understanding of the complex magnetic field behavior observed in these objects, with possible implications for a broader class of dynamos in astrophysical objects ~\citep{Jones2011}. One of the major drawbacks of such simulations is that there is order of magnitude disagreement between the natural and the numerically accessible values of several control parameters. For instance, in numerical simulations of the geodynamo, the Ekman number -- a nondimensional measure of the importance of viscous effects as compared to the Coriolis effects -- is usually five to ten orders of magnitude larger than the expected realistic values. 

One way to tackle this disparity is to infer asymptotic scaling laws from a sufficient number of numerical results. Such numerical scaling laws can then be extrapolated to realistic parameter regimes and compared with the observational data. \cite{Christensen2006} (hereafter ``CA6") used a battery of numerical simulation results to derive scaling relations for heat transfer, convective velocity, and magnetic field strength. Their scaling relations hold over several orders of magnitude of the relevant control parameter. Using these scaling relations, CA6 predicted magnetic field strengths inside the Earth's and Jupiter's core and found reasonable agreement with observationally constrained values. Later, \cite{Christensen2009} showed that these scaling laws are also in good agreement with magnetic fields observed in fast rotating low-mass stars. \cite{Takahashi2008} and \cite{Aubert2009} independently reinforced the scaling laws put forth by CA6. \cite{Olson2006} derived scaling laws specifically for  the dipole moment of planetary dynamos which show an order-of-magnitude agreement with the observed dipole moments of solar system planets. \cite{Christensen2010} reviews earlier scaling laws for planetary magnetic field based on heuristic arguments and compares them with the numerically established scaling relations. 

The mechanical boundary conditions may play an important role in the dynamo mechanism. Dynamos which operate in planets with a solid mantle are usually modeled with rigid boundaries~\citep{Kageyama1995, Glatzmaier1995a, Glatzmaier1995b}. While a true free surface is difficult to model, a free-slip condition (i.e. assuming zero shear stress at an undeformable spherical boundary) is a much better approximation than a no-slip condition for the surface of gas- and ice-giant planets or stellar convection zones. Rigid boundaries are associated with viscous (Ekman) boundary layers, which have a damping influence on the development of strong axisymmetric flows that are found in free surface flows. \cite{Kuang1997, Kuang1999} argued that even for the geodynamo a free-slip condition may be a better choice, because the Ekman layers in the models are much thicker than the very thin layers in the Earth's core. \cite{Aubert2001} demonstrated in rotating liquid Gallium experiments that at small  Prandtl numbers -- the  ratio of kinematic viscosity to thermal diffusivity -- flows with rigid boundaries show features similar to those with free-slip boundaries, i.e. dominant zonal flows.  \cite{Miyagoshi2010} also found strong zonal flows in low Ekman number ($\approx10^{-7}$) rigid boundary geodynamo simulations. Hence, studies of dynamos with free-slip boundaries have wide ranging applications. Following these arguments, researchers have modelled Uranus' and Neptune's multipolar magnetic field~\citep{Stanley2004, Stanley2006}, Mercury's weak magnetic field~\citep{Stanley2005}, and Saturn's unusually axisymmetric magnetic field~\citep{Stanley2010a} using free-slip boundaries. 

Dynamos with free-slip boundaries also exhibit bistability: the morphology of the dynamo generated magnetic field depends on the initial magnetic field configuration~\citep{Simitev2009, Simitev2012, Sasaki2011, Schrinner2012, Gastine2012dyn}. Recently, \cite{Gastine2013} found evidence of bistability in M dwarfs. Grote and co-workers \citep{Grote1999, Grote2000a, Grote2000b} have employed free-slip boundaries in their dynamo models, and found a wider spectrum of magnetic field geometries than what has been reported for rigid boundaries. 

A direct comparison of the effects of different mechanical boundary conditions on the dynamo has rarely been made. \cite{Christensen1999} reported results for both kinds of boundary conditions for a limited number of cases and found that the large scale magnetic field is similar for both cases. Recently, \cite{Schrinner2012} have analyzed many dynamo simulations with rigid, free-slip and mixed (rigid at inner and free-slip at outer boundary) boundary conditions and report a difference in magnetic field amplitude of dipolar and multipolar dynamos. Following this study, we specifically focus here on deriving scaling properties for heat transport, velocity, and magnetic field strength. We compare our findings  with earlier rigid boundary systems for which extensive modeling results are available in a broad range of control parameters (CA6). This exercise helps us in isolating the effect of mechanical boundary condition.

\section{\label{sec:dyn_mod}Dynamo model}
\subsection{MHD equations}
Our numerical set-up consists of a spherical shell which rotates along the $\hat{z}$-axis and which is bounded by inner radius $r_i$ and outer radius $r_o$. The aspect ratio $r_i/r_o$ is 0.35. A linear variation of gravity with radius is assumed. Following CA6, we non-dimensionalize  the magnetohydrodynamic (MHD) equations by using the shell thickness $r_o-r_i=D$ as the reference length scale and $1/\Omega$, where $\Omega$ is the rotation rate, as the time unit. The magnetic field $\mathbf{B}$ is scaled by $\Omega D \sqrt{\rho\mu}$, where $\rho$ is the constant fluid density and $\mu$ is the magnetic permeability. Note that all of the above non-dimensional scales are free from any diffusion parameters.  The temporal evolution of velocity $\mathbf{u}$, temperature $T$, and magnetic field $\mathbf{B}$ is governed by the MHD equations under the Boussinesq approximation
\begin{gather}
\frac{\partial\mathbf{u}}{\partial t}+\mathbf{u\cdot\nabla\mathbf{u}}+2\hat{z}\times\mathbf{u}+\nabla P = \nonumber \\
Ra^{*}\frac{\mathbf{r}T}{r_{o}}+(\nabla\times\mathbf{B})\times\mathbf{B}+E\nabla^{2}\mathbf{u}, \label{eq:MHD_vel} \\
\frac{\partial T}{\partial t}+\mathbf{u\cdot\nabla}T  = \frac{E}{Pr}\nabla^{2}T, \\
\frac{\partial\mathbf{B}}{\partial t}  =  \nabla\times(\mathbf{u}\times\mathbf{B})+\frac{E}{Pm}\nabla^{2}\mathbf{B}, \label{eq:MHD_mag}\\
\nabla \cdot \mathbf{u}  =  0,   \\
\nabla \cdot \mathbf{B}  =  0.  \label{eq:div_B_0}
\end{gather}
This system of equations is governed by several nondimensional control-parameters: Ekman number $E=\nu/\Omega d^2$, $\nu$ being the fluid viscosity; the modified Rayleigh number $Ra^{*}=\alpha g_o \Delta T/\Omega^2 D$, where $g_o$ is gravity at the outer boundary and $\alpha$ is the thermal expansivity; magnetic Prandtl number $Pm=\nu/\eta$, $\eta$ being the magnetic diffusivity; Prandtl number $Pr=\nu/\kappa$, $\kappa$ being the thermal conductivity. $Ra^{*}$ is related to the conventional Rayleigh number $Ra$ through $Ra^{*}=RaE^2/Pr$.

We assume free-slip mechanical boundaries at both inner and outer radius. The magnetic field matches a potential field at both boundaries. A fixed temperature contrast $\Delta T$ is maintained between the top and the bottom.

\subsection{Numerical method}
Equations (\ref{eq:MHD_vel}-\ref{eq:div_B_0}) are numerically solved using the MagIC code~\citep{Wicht2002}. Velocity and magnetic field are first separated into toroidal and poloidal components as
\begin{gather}
\mathbf{u}=\nabla\times u_T\hat{r} + \nabla\times\nabla\times u_P\hat{r}, \nonumber \\
\mathbf{B}=\nabla\times B_T\hat{r} + \nabla\times\nabla\times B_P\hat{r}. \nonumber
\end{gather}
The scalar potentials $u_{T,P}$ and $B_{T,P}$, along with temperature $T$ and pressure $P$, are further expanded using spherical harmonics in the $\theta$ and $\phi$ directions and the Chebyshev polynomials in the radial direction. $N_r$ and $l_{max}$ are the maximum degree of the Chebyshev polynomials and the spherical harmonic functions used in this expansion. For all the simulations considered here, $41\leq N_r \leq 73$ and $64\leq l_{max}\leq 170$. The simulations are run for at least one magnetic diffusion time ($D^2/\eta$) to ensure a statistically stationary state.

\subsection{Diagnostic parameters}
We employ several diagnostic parameters to analyze our simulations results. The Rossby number $Ro$ is the volume averaged non-dimensional rms velocity. Following CA6, we also introduce the local Rossby number $Ro_l$ which is a more appropriate measure than $Ro$ to characterize the ratio of the inertial and the Coriolis forces. A typical flow length scale can be calculated based on the mean spherical harmonic degree $l$
\begin{gather}
\bar{l}_u=\underset{l}{\sum} l \frac{\langle \mathbf{u}_l \cdot \mathbf{u}_l \rangle}{\langle \mathbf{u} \cdot \mathbf{u} \rangle}, \nonumber
\end{gather}
where $\langle...\rangle$ denotes time average and $\mathbf{u}_l$ is velocity component at degree $l$. The local Rossby number is then defined as $Ro_l=Ro\,\pi/\bar{l}_u$.

The volume averaged non-dimensional rms magnetic field strength is called Lorentz number $Lo$. The field geometry at the outer boundary surface is characterized by its dipolarity $f_{dip}$. It is defined as the ratio of the magnetic energy of the dipole to the total magnetic energy at the outer boundary surface. 

The Nusselt number $Nu$ is a ratio of total heat transported from the inner shell to the outer shell to the conducted heat. It is expected that different diffusivities play a minimal role in determining the large scale properties of the dynamo systems. This motivated CA6 to define a modified Nusselt number $Nu^*$ which does not involve $\kappa$. $Nu^*$ is related to the conventional Nusselt number $Nu$ via $Nu^{*}=(Nu-1)E/Pr$. In addition, the heat flux from surfaces of astrophysical objects is a much more meaningful and accessible quantity than the temperature difference between the inner and outer boundary of the convection zone. CA6 defined a heat flux based Rayleigh number $Ra^*_Q$ which incorporates the advected heat flux rather than the temperature contrast. $Ra^*_Q$ is related to $Ra$ through $Ra^*_Q = Ra(Nu-1)E^3/Pr^2$.

The reported numerical values of the diagnostic parameters $Ro$, $Ro_l$, $Lo$, $Nu^*$, and $f_{dip}$ are time averaged values excluding the initial transients.

\section{\label{sec:results}Results}
We have built up a data set of 57 dynamo simulations with free-slip boundaries: 40 cases by us (see Table~\ref{tab:table2}) and 17 cases adopted from~\cite{Schrinner2012}. All of the cases in this data set have the same physical set-up as described above. Also, following CA6, we report and analyze only those dynamo simulations which have $Nu>2$ to ensure a vigorous enough convection that fills the full volume of the spherical shell.

\subsection{\label{sec:bis}Bistability}
Bistability is a phenomenon in which a system shows two different dynamo solutions, i.e. dipolar and multipolar, for the same set of control parameters, but with different initial conditions for the magnetic field. For dynamos with free-slip boundaries, \cite{Simitev2009} found two distinct dipolar and multipolar dynamo branches, at least in some parameter range. \cite{Sasaki2011}, \cite{Schrinner2012}, and  \cite{Gastine2012dyn} have also observed bistability in their dynamo simulations with free-slip boundaries. Dynamos in box geometry with periodic boundaries also show bistability~\citep{Yadav2012}; in fact, even tristable and quadstable solutions were observed in such simulations. Bistable dynamo solutions have rarely been observed in dynamos with rigid boundaries, e.g. CA6 reported a single case in which they found bistable states. In dynamos with rigid boundaries, the dipolar branch collapses around $Ro_{l}\approx 0.1$, which CA6 argue is due to inertial forces dominating over Coriolis force at larger $Ro_{l}$, while~\cite{Soderlund2012} argue that the dipole collapse is due to helicity degradation related to the competition of inertial and viscous forces. \cite{Soderlund2012} also hypothesize that current numerical planetary dynamo models are viscously controlled.

\begin{figure}
\begin{center}
\includegraphics[scale=0.47]{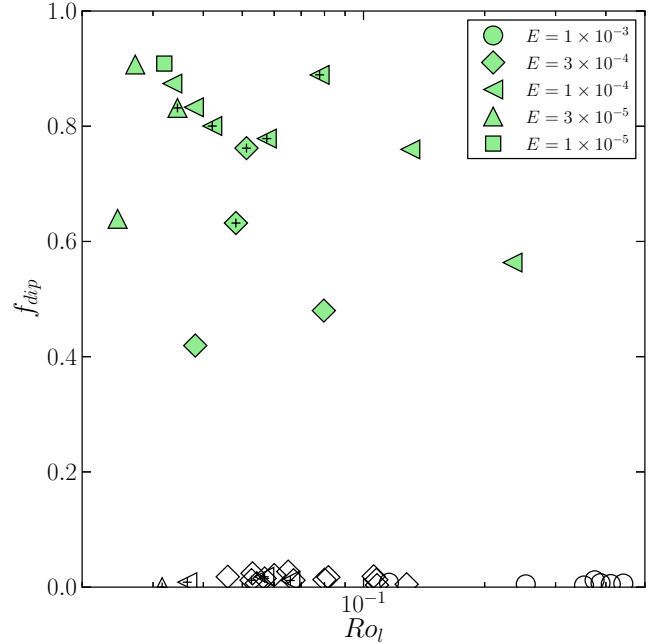}
\caption{Dipolarity at the outer boundary versus the local Rossby number. The data points carrying ``+" marker are bistable states. Filled (empty) symbols are dipolar (multipolar) dynamos. The symbol shapes represent the Ekman number and the corresponding value is given in the upper-right-corner box.}
\label{fig:fdip_Rol}
\end{center}
\end{figure}

We plot in Fig.~\ref{fig:fdip_Rol} the dipolarity $f_{dip}$ at the outer surface ($r=r_o$) versus the local Rossby number $Ro_l$ (for our simulations). In the dipolar branch (filled data points) the magnetic field is dominated by the dipole ($f_{dip}>0.4$) and in the multipolar branch (empty data points) the dipole is much weaker ($f_{dip}<0.05$). The dipole branch is limited to cases with  $Ro_l\lesssim0.2$. However, the multipolar branch exists for a broad range of $Ro_l$. The highly supercritical dipolar case at $Ro_l\approx0.2$ ($E=1\times10^{-4}$, $Pm=0.5$ and $Ra = 4\times 10^7$) was run for three magnetic diffusion times without an indication of a dipole collapse, although we can not exclude that the field could change to multipolar in the long run. Dipolar dynamos at such high $Ro_l$ have not been reported yet. Simulations that settle down to different dynamo states depending on the initial magnetic field are marked with a ``+" in Fig.~\ref{fig:fdip_Rol} (difficult to discern on the multipolar branch due of clustering). Note that other dipolar dynamos could show bistability but we did not explore all of our dipolar cases for such behavior.

For rigid boundary dynamos, CA6 reported that multipolar dynamo solutions are not observed for $Ro_l<0.1$. But results from \cite{Schrinner2012} and our findings suggest that this is not the case for free-slip boundaries. We also found a few dynamo solutions, which have $Ro_l<0.1$, but settle to a multipolar solution despite having initial dipolar magnetic field. For example, we only found  multipolar solutions for $E=3\times10^{-4}$ and $Pm\le1.5$. This demonstrates that, depending on the control parameters, only the multipolar dynamo branch can be stable in some situations. This is in agreement with earlier results \citep{Simitev2009, Simitev2012} which showed that bistability is in fact a function of $P$, $Pm$ and $E$. They used volumetric heating while we use a fixed temperature contrast to drive  convection. The bistable behavior we observed generalizes their findings.

One of the characteristic features of rotating spherical shell convection with free-slip boundaries is the development of strong axisymmetric zonal flows. In rigid boundary systems, the Reynolds stresses, which arise due to a statistical correlation between the radial and azimuthal flow component (in cylindrical co-ordinates), are balanced by the bulk viscosity and the Ekman layer friction near the outer boundary. In the case of free-slip boundaries, the Ekman layers are absent and zonal flows can thus saturate at much higher vigor. In dynamo models, Maxwell stresses also affect the zonal flows; these stresses are potentially higher in dipolar dynamos that have higher magnetic field strength than multipolar ones at the same control parameter values~\citep{Browning2008}. One argument for the essential role of zonal flows for bistability is that an initial dipolar magnetic field inhibits the growth of zonal flows via Maxwell stresses. In the case of a multipolar initial condition, strong zonal flows can develop, which in turn suppress the development of dipolar magnetic fields.  This mechanism allows multipolar magnetic fields even for $Ro_l$ smaller than 0.1. 

\begin{figure}
\begin{center}
\includegraphics[scale=0.28]{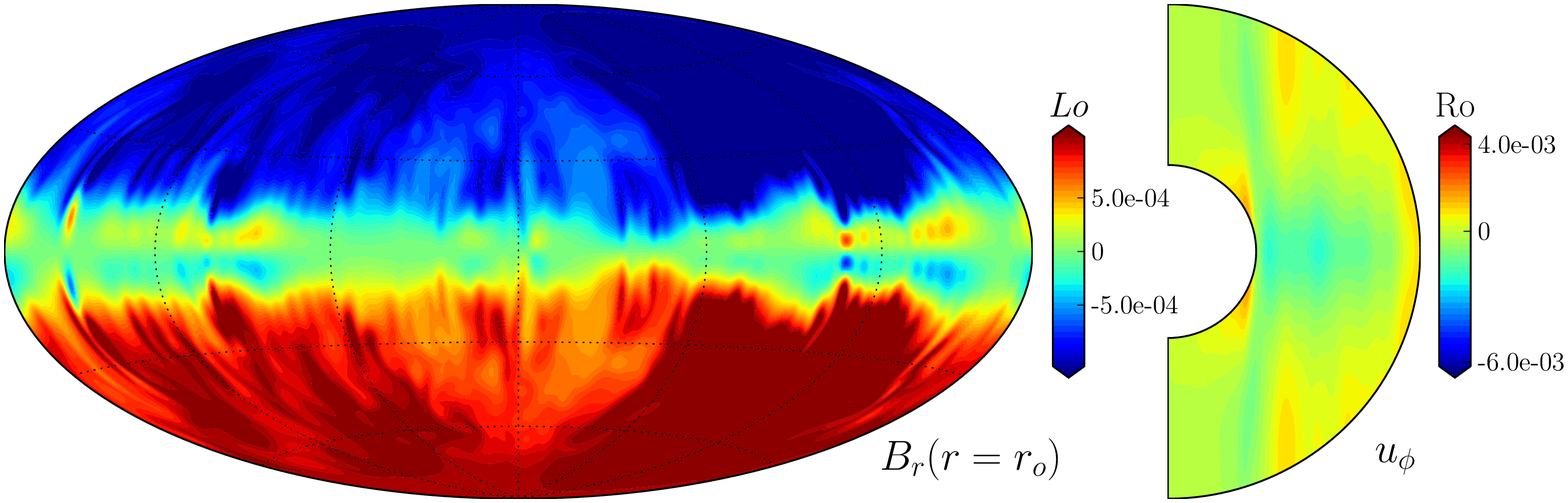} \\
\includegraphics[scale=0.28]{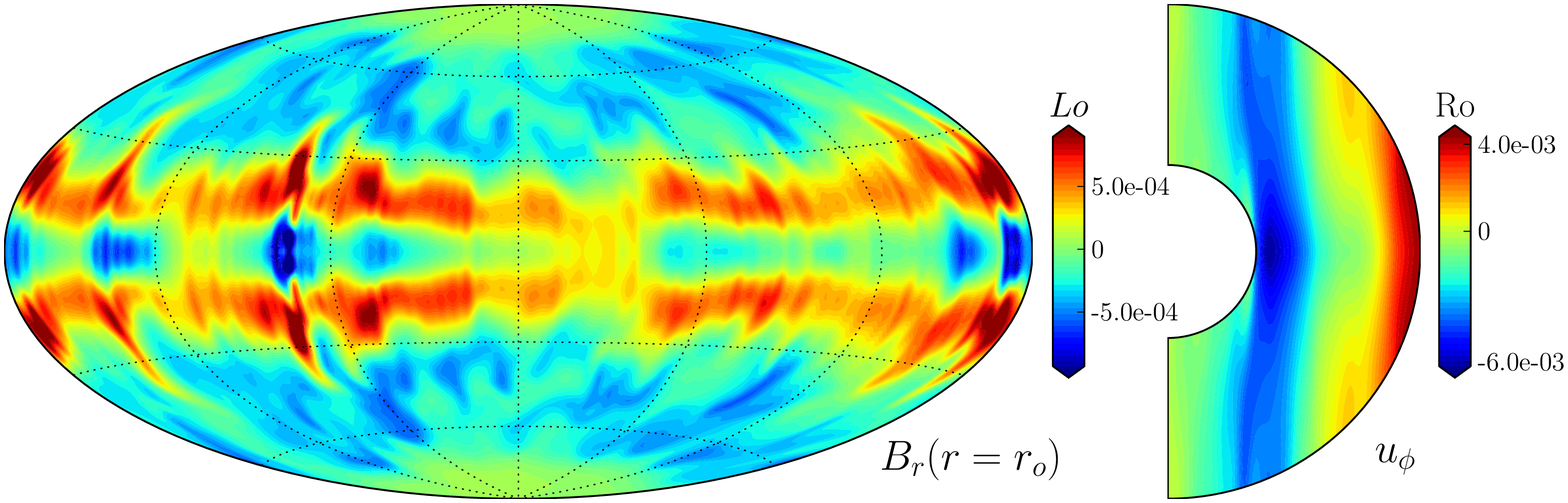} 
\caption{A snapshot of the non dimensional radial magnetic field at outer boundary and zonal flow (azimuthally averaged) in a meridional section for a dipolar (top row) and a multipolar (bottom row) dynamo. These states are obtained at $E=3\times10^{-5}$, $Pm=0.5$, and $Ra = 3\times10^{7}$. The radial magnetic field is truncated at 50\% of the maximum in order to highlight the magnetic field structures.}
\label{fig:zonal_flow}
\end{center}
\end{figure}

The zonal flow structure of a bistable state is shown in Fig.~\ref{fig:zonal_flow}. It portrays a weak thermal wind driven zonal flow in the dipolar case and a nearly three times stronger and more geostrophic zonal flow in the multipolar case. Note that the magnetic field of the multipolar solution has a quadrupolar symmetry, but this is not generally the case. In particular in strongly driven cases the magnetic field has a smaller length scale and does not have any preferred symmetry. \cite{Aubert2005} found that the thermal wind driven zonal flow topology is in agreement with Ferraro's law of co-rotation~\citep{Ferraro1937}, i.e. the shearing of the axisymmetric poloidal magnetic field by the zonal flow is minimal. The non-geostrophy of the flow in case of the dipolar dynamo emphasizes that the zonal flow quenching by the Lorentz force is rather large in dipolar dynamos as compared to that in the multipolar dynamo cases, as also pointed out before by \cite{Schrinner2012}.

\subsection{Nusselt number scaling}
\begin{figure}
\begin{center}
\includegraphics[scale=0.47]{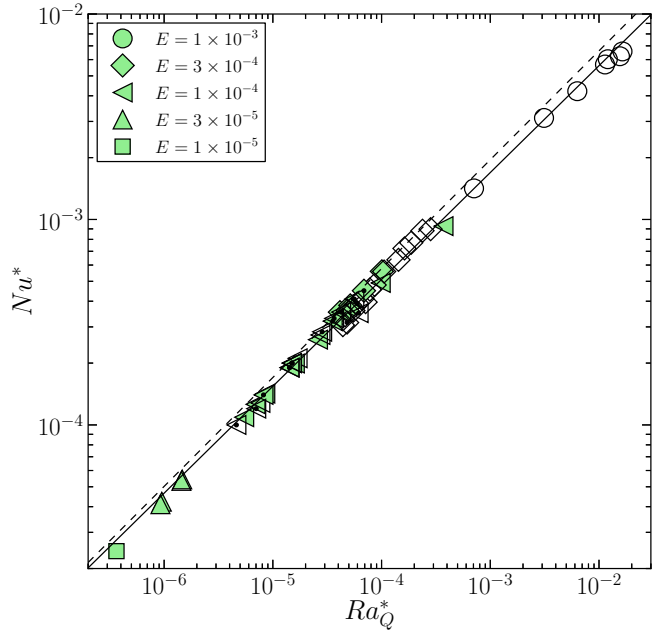}
\caption{Modified Nusselt number versus heat flux based Rayleigh number. The data symbols carrying a thick black dot are adopted from \cite{Schrinner2012}. The solid-line is a line-fit to data and the dashed-line is the scaling reported by CA6 for dipolar rigid boundary dynamos.}
\label{fig:Nu_RaQ}
\end{center}
\end{figure}

Figure~\ref{fig:Nu_RaQ} shows that the modified Nusselt number $Nu^*$ scales very well with the flux-based Rayleigh number $Ra^*_Q$ in the same way irrespective of whether the dynamo is dipolar (filled symbols) or multipolar (empty symbols). A best-fit line to this data set reveals a relation $Nu^*=0.061\,{Ra^*_Q}^{0.52}$, which agrees well with the scaling $Nu^*=0.076\,{Ra^*_Q}^{0.53}$  (dashed line) found by CA6 for dipolar dynamos with rigid boundaries. For hydrodynamic convection with free-slip boundaries, \cite{Christensen2002} suggested a possible asymptotic scaling $Nu^*=0.077\,{Ra^*_Q}^{5/9}$ in the limiting case of  $E\rightarrow 0$. In relatively thinner shells ($r_i/r_o=0.6$), \cite{Gastine2012} report $Nu^*=0.086\,({\langle Ra^*_Q \rangle_{\rho}})^{0.53}$, where $\langle ... \rangle_{\rho}$ designates mass-averaged quantities, for density stratified anelastic hydrodynamic convection simulations with free-slip boundaries. All these scaling relations are very close to each other, which suggests that magnetic field, mechanical boundary conditions or density stratification  have no substantial effect on the scaling behavior of heat transport in rotating spherical shell convection. However, we note that for larger values of the Rossby number than the ones considered here, in a regime where inertia dominates over the Coriolis force, the power-law scaling between $Nu^*$ and $Ra^{*}_Q$  breaks down \citep{King2009, King2010, Schmitz2010}.

\subsection{\label{sec:ros}Rossby number scaling}
\begin{figure}
\begin{center}
\includegraphics[scale=0.47]{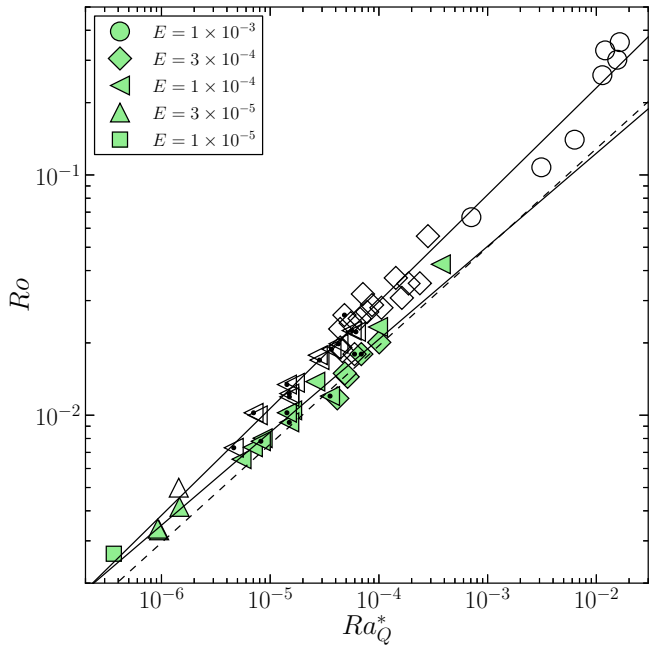}
\caption{Rossby number versus the heat flux based Rayleigh number. The two solid lines are best-fitting lines to dipolar and multipolar dynamos. The dashed-line represent the scaling reported by CA6 for dipolar rigid boundary dynamos.}
\label{fig:Ro_RaQ}
\end{center}
\end{figure}

In Fig.~\ref{fig:Ro_RaQ} we plot $Ro$ as a function of $Ra^*_Q$. The data points are slightly more scattered as compared to Fig.~\ref{fig:Nu_RaQ}, but a consistent scaling is nonetheless evident. Moreover, a somewhat different scaling for dipolar and multipolar dynamos is visible as demonstrated by the two different solid lines. These lines are $Ro=0.73\,{Ra^*_Q}^{0.39}$ (dipolar) and $Ro=1.79\,{Ra^*_Q}^{0.44}$ (multipolar). The scaling reported by CA6 for dipolar rigid boundary dynamos is $Ro=0.85\,{Ra^*_Q}^{0.41}$ (dashed-line), which agrees with our dipolar dynamo scaling.

\begin{figure}
\begin{center}
\includegraphics[scale=0.47]{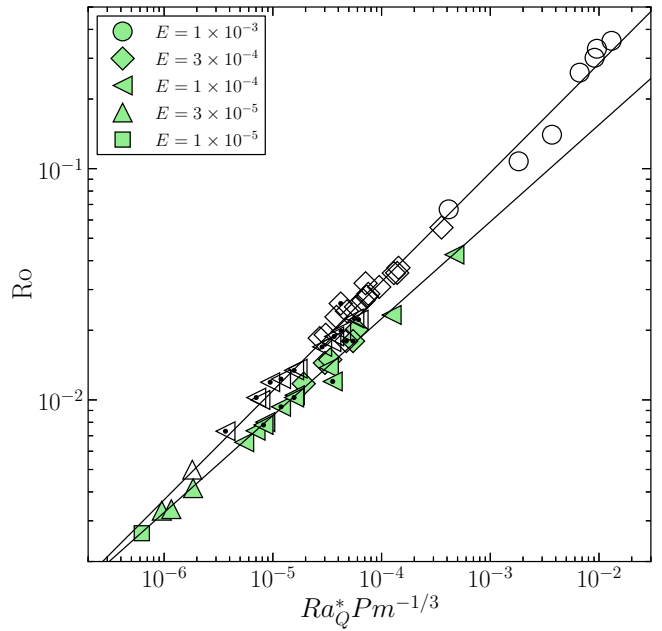}
\caption{Rossby number scaling incorporating a $Pm$ dependence.}
\label{fig:Ro_RaQPm}
\end{center}
\end{figure}

Similar to CA6, the scatter in Fig.~\ref{fig:Ro_RaQ} can be reduced to some extent by assuming an additional $Pm$ dependence. A two-parameter least-square-optimized fit provides $Ro=0.99{Ra^*_Q}^{0.41}Pm^{-0.1}$ (dipolar) and $Ro=2.44\,{Ra^*_Q}^{0.47}Pm^{-0.14}$ (multipolar). This optimization reduces the standard error by almost 48\% in the dipolar scaling and 20\% in the multipolar scaling (see Table~\ref{tab:table1}). We also considered the Ekman number as additional parameter for improving the fit, but, as observed by CA6, the resulting exponents are rather small as compared to $Ro$ and $Pm$ exponents. Hence, we discard a dependence on $E$ in our scaling analysis. The $Pm$ exponent is small and appears to depend on the nature of the magnetic field. The latter could be an artifact of the relatively small size of the data set, especially for dipolar dynamos. In fact, similar to CA6, assuming a scaling of the form $Ro\propto{(Ra_Q^* Pm^{-1/3})}^{\alpha}$, a good fit is obtained for both dipolar and multipolar cases, with somewhat different values for $\alpha$ (see Table~\ref{tab:table1}): this form is shown in Fig.~\ref{fig:Ro_RaQPm}. 

\begin{figure}
\begin{center}
\includegraphics[scale=0.47]{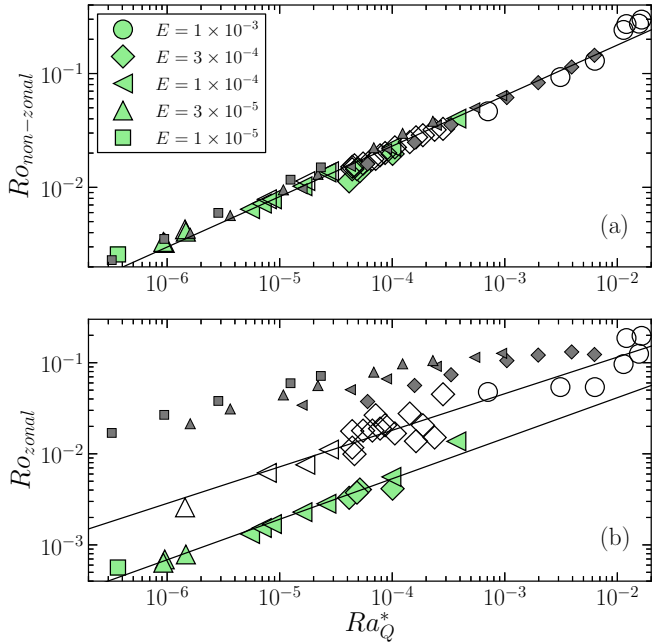}
\caption{Non-zonal Rossby number in (a) and zonal Rossby number in (b) versus the heat flux based Rayleigh number. The gray filled data points are simulation results ($Nu>2$) of hydrodynamic free-slip convection from \cite{Christensen2002}.}
\label{fig:zonal_and-non-zonal_Ro}
\end{center}
\end{figure}

As described in Sec.~\ref{sec:bis}, zonal flows are stronger in rotating convective shells with free-slip mechanical boundaries. In the corresponding MHD systems, dynamos with multipolar magnetic fields will have stronger zonal flows as compared to those with dipolar magnetic fields. This effect is visible in Fig.~\ref{fig:zonal_and-non-zonal_Ro} whose top panel shows $Ro_{non-zonal}$ versus $Ra^*_Q$, and the bottom panel shows $Ro_{zonal}$ versus  $Ra^*_Q$. $Ro_{non-zonal}$ and $Ro_{zonal}$ are calculated by considering the rms velocity excluding the axisymmetric zonal-flow component and the rms velocity of only the axisymmetric zonal flow, respectively. The scaling in Fig.~\ref{fig:zonal_and-non-zonal_Ro}a is $Ro_{non-zonal}=1.37\,{Ra^*_Q}^{0.44}$ and in Fig.~\ref{fig:zonal_and-non-zonal_Ro}b is $Ro_{zonal}=0.32\,{Ra^*_Q}^{0.44}$ (dipolar) and $Ro_{zonal}=0.73\,{Ra^*_Q}^{0.4}$ (multipolar) We also considered $Pm$ as an additional scaling parameter (see Table~\ref{tab:table1}). The resulting scaling is marginally better, but the improvement is not as remarkable as it is in Fig.~\ref{fig:Ro_RaQPm}. 

As illustrated in Fig.~\ref{fig:zonal_and-non-zonal_Ro}a the non-zonal flow component is unaffected by magnetic field geometry as both dipolar and multipolar dynamos follow the same $Ro_{non-zonal}$ scaling. To further investigate the effect of magnetic field on the flow, Fig.~\ref{fig:zonal_and-non-zonal_Ro} also incorporates results (gray filled symbols) of hydrodynamic convection in spherical shells with free-slip boundaries from \cite{Christensen2002}. This reveals that  magnetic field itself does not affect the scaling behavior of $Ro_{non-zonal}$. Coupled with our earlier observation that $Nu^*$ scaling is effectively same in hydrodynamic and magnetohydrodynamic convection in spherical shells, we can conjecture that the scaling of the flow component which is responsible for heat transfer is unaffected by the presence of magnetic field.

Unlike the non-zonal Rossby number, the zonal Rossby number of dipolar dynamos  is consistently lower than that of corresponding multipolar dynamos (Fig.~\ref{fig:zonal_and-non-zonal_Ro}b). This difference in zonal flows  explains the offset in the Rossby number scaling in dipolar and multipolar dynamos seen in Fig.~\ref{fig:Ro_RaQ} and~\ref{fig:Ro_RaQPm}. Note that as compared to the dipolar branch the scatter in the multipolar branch of Fig.~\ref{fig:zonal_and-non-zonal_Ro}b is large. This could be due to the fact that unlike the dipolar branch the multipolar branch is a blend of dynamos which have quadrupolar, octupolar, and sometimes even higher order modes as the most dominating magnetic mode. Since the Maxwell stresses are dependent on the magnetic field geometry, the zonal flows saturate at many different levels. The hydrodynamic zonal flow is consistently higher than both dipolar and multipolar cases. For dipolar dynamos with rigid boundaries, \cite{Aubert2005} argued that Lorentz forces are essential to saturate the zonal flow and bring it into a thermal wind balance, rather than the boundary friction. In purely hydrodynamic cases with free-slip boundaries, it must be viscous friction in the bulk volume that limits the amplitude of the zonal flow. Even for our multipolar dynamos, the zonal flow amplitude is smaller than in the hydrodynamic cases (Fig.~\ref{fig:zonal_and-non-zonal_Ro}b), which indicates that Lorentz forces play an important role for saturating the flow. For the dipolar dynamos with stronger magnetic fields the damping effect is more pronounced.

\subsection{Magnetic field scaling}
\begin{figure}
\begin{center}
\includegraphics[scale=0.47]{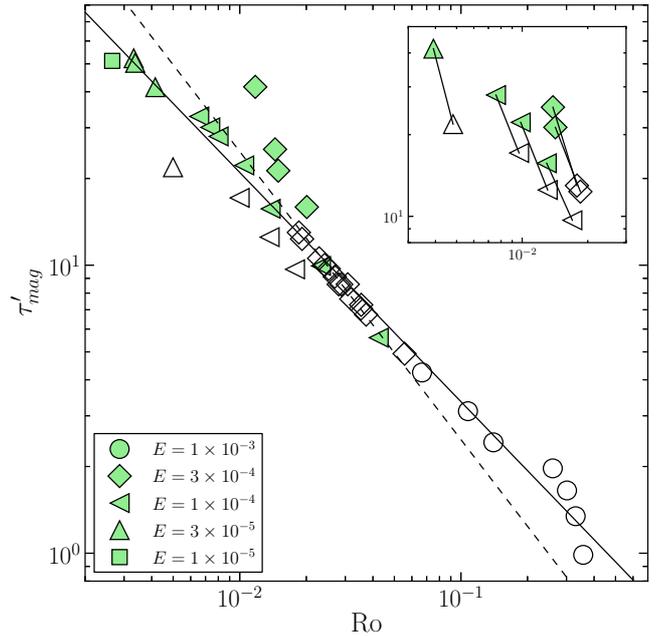}
\caption{Ohmic dissipation time versus the Rossby number. Solid line represents a best fit line, while the dashed line represents $\tau'_{mag}\propto1/Ro$. The inset-figure contains bistable pairs connected by solid lines.}
\label{fig:tmag_Ro_combo}
\end{center}
\end{figure}

The Ohmic dissipation time $\tau_{mag}$, which is the ratio of magnetic energy and Ohmic dissipation, is a function of the typical length scale of the magnetic field. As the magnetic Reynolds number is increased, the magnetic field becomes smaller scaled, and, since small scales are associated with faster time scales, the Ohmic dissipation time scale decreases. This qualitative argument was verified by \cite{Christensen2004} in rigid boundary spherical shell dynamos. They showed that $\tau_{mag}$ (normalized with magnetic diffusion time) is approximately inversely proportional to the magnetic Reynolds number $Rm$. When $\tau'_{mag}$ is the Ohmic dissipation time expressed in units of rotation period of the spherical shell, this inverse relation translates to $\tau'_{mag}\propto1/Ro$.  In Fig.~\ref{fig:tmag_Ro_combo} we plot $\tau'_{mag}$ versus $Ro$. A best-fit line to this data set suggests $\tau'_{mag}\propto1/Ro^{0.8}$. Since the scatter in Fig.~\ref{fig:tmag_Ro_combo} is substantial, the difference between the exponents -0.8 and -1 may not be very significant. \cite{Christensen2010} have discussed a more complex scaling for $\tau_{mag}$ and report a marginal improvement in the quality of the fit. Although the inset figure shows a small decrease in $\tau'_{mag}$ for bistable states when the magnetic field is multipolar, the scalings for dipolar and multipolar dynamos appear to follow the same trend. Moreover, if we plot $\tau'_{mag}$ versus $Ro_{zonal}$ or $Ro_{non-zonal}$ (not shown), then the scatter is increased as compared to Fig.~\ref{fig:tmag_Ro_combo}. It highlights that the important parameter in the context of ohmic dissipation is the {\em total} Rossby number, which incorporates the zonal-flow contribution.

\begin{figure}
\begin{center}
\includegraphics[scale=0.47]{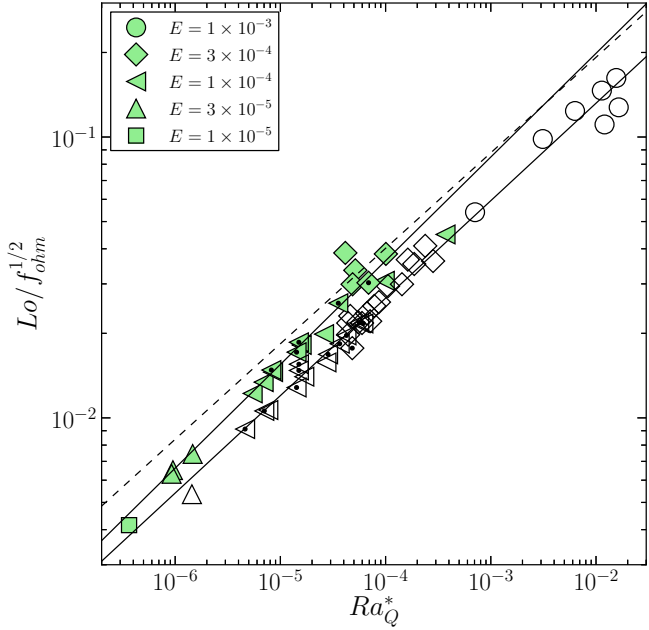}
\caption{Lorentz number versus the heat flux based Rayleigh number. The two solid lines are best-fitting lines to dipolar and multipolar dynamos. The dashed-line represent the scaling reported by CA6 for dipolar rigid boundary dynamos.}
\label{fig:Lofohm_RaQ}
\end{center}
\end{figure}

\begin{figure}
\begin{center}
\includegraphics[scale=0.47]{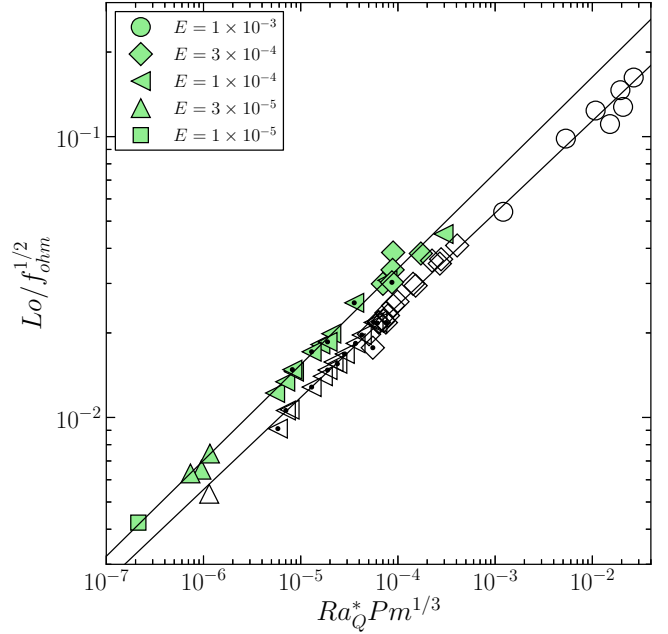}
\caption{Lorentz number scaling incorporating a $Pm$ dependence.}
\label{fig:Lofohm_RaQPm}
\end{center}
\end{figure}

CA6 argue that the magnetic field strength might be determined by the power available to balance the Ohmic dissipation. Following this argument, the Lorentz number should be accordingly corrected by the Ohmic fraction $f_{ohm}$ which is the ratio of Ohmic dissipation and the power generated via buoyancy forces. In Fig.~\ref{fig:Lofohm_RaQ}, we plot the corrected Lorentz number versus the flux-based Rayleigh number $Ra^{*}_Q$, which is a non-dimensional measure for the power generated by the action of buoyancy forces (CA6). A best fit is obtained by $Lo/f^{1/2}_{ohm}=1.08\,{Ra^*_Q}^{0.37}$   (dipolar) and $Lo/f^{1/2}_{ohm}=0.65\,{Ra^*_Q}^{0.35}$ (multipolar). The dipolar scaling in Fig.~\ref{fig:Lofohm_RaQ} is close to the rigid boundary dipolar scaling $Lo/f^{1/2}_{ohm}=0.92\,{Ra^*_Q}^{0.34}$ (dashed line) reported by CA6. Furthermore, a two-parameter optimized fit for dipolar dynamos is $Lo/f^{1/2}_{ohm}=0.72\,{Ra^*_Q}^{0.33}Pm^{0.14}$ and for multipolar dynamos is $Lo/f^{1/2}_{ohm}=0.51\,{Ra^*_Q}^{0.33}Pm^{0.11}$. The inclusion of $Pm$ reduces the standard error by almost 67\% (dipolar) and 39\% (multipolar).  Again, assuming a simplified form $Lo/f^{1/2}_{ohm}\propto ({Ra^*_Q}Pm^{1/3})^{\beta}$, the quality of the fit is hardly reduced (Table~\ref{tab:table1}, Fig.~\ref{fig:Lofohm_RaQPm}). A cursory inspection of Figs.~\ref{fig:Lofohm_RaQ} and \ref{fig:Lofohm_RaQPm} suggests that the dipolar and multipolar scalings are almost the same except for an offset in the pre-factor by  $\approx8/5$. 

An inverse relation of $\tau'_{mag}$ and $Ro$ translates to $Lo/f^{1/2}_{ohm}\propto\sqrt{Ra^*_Q/Ro}$. If we now substitute the $Ro$ scaling from Fig.~\ref{fig:Ro_RaQ} in the previous relation, then an offset of $\approx8/5$ is indeed expected for dipolar and multipolar scaling. The $\tau'_{mag}\propto1/Ro$ argument therefore supports the offset in the scaling observed in Fig.~\ref{fig:Lofohm_RaQ} to a good extent. \cite{Schrinner2012} have also reported a similar shift in their free-slip dynamo simulations; they qualitatively argue that the offset in the scaling is due to decrease in the $f_{ohm}$ in multipolar dynamo cases.  In the case of rigid boundaries, \cite{Christensen2010} also reported a smaller scaling pre-factor of the Lorentz number for multipolar dynamos compared to dipolar ones. Our inspection of data from earlier spherical shell dynamos with rigid boundaries reveals that $Ro$ for both dipolar and multipolar dynamos follows same scaling relation, unlike what we observed. Clearly, more analysis is required to conclusively demonstrate the reason for such an offset.

\begin{table}
\centering
\caption{\label{tab:table1}The various scaling laws inferred from our study, along with the cross-correlation coefficient $R$ and the Standard error (standard deviation divided by square-root of number of data points).}
\begin{tabular}{lcc}
\toprule
Scaling &$R$ &Standard-error \\
\midrule
$Nu^*=0.061\,{Ra^*_Q}^{0.52}$ 			&0.9987 	&0.0036 \\
$Ro_{non-zonal}=1.37\,{Ra^*_Q}^{0.44}$		&0.9924		&0.0089 \\
$Ro_{non-zonal}=1.99\,({Ra^*_Q}Pm^{-1/3})^{0.47}$		&0.9937	&0.0086 \\
\midrule
Dipolar& &\\
$Ro=0.73\,{Ra^*_Q}^{0.39}$  			&0.9903		&0.0128 \\
$Ro=0.99{Ra^*_Q}^{0.41}Pm^{-0.1}$  		&0.9976		&0.0067 \\
$Ro=1.07\,({Ra^*_Q}Pm^{-1/3})^{0.42}$	  	&0.9966		&0.0082 \\
$Ro_{zonal}=0.32\,{Ra^*_Q}^{0.44}$		&0.9873		&0.0198 \\
$Ro_{zonal}=0.47\,({Ra^*_Q}Pm^{-1/3})^{0.48}$		&0.9896		&0.0192\\
$Lo/f^{1/2}_{ohm}=1.08\,{Ra^*_Q}^{0.37}$ 		&0.9820		&0.0167 \\
$Lo/f^{1/2}_{ohm}=0.71\,{Ra^*_Q}^{0.33}Pm^{0.14}$	&0.9972		&0.0059 \\
$Lo/f^{1/2}_{ohm}=0.78\,({Ra^*_Q}Pm^{1/3})^{0.34}$	&0.9967		&0.0066 \\
\midrule
Multipolar& &\\
$Ro=1.79\,{Ra^*_Q}^{0.44}$ 				&0.9916		&0.0098 \\
$Ro=2.44\,{Ra^*_Q}^{0.47}Pm^{-0.14}$ 			&0.9954		&0.0078 \\
$Ro=2.49\,({Ra^*_Q}Pm^{-1/3})^{0.47}$ 	 		&0.9952		&0.0078 \\
$Ro_{zonal}=0.73\,{Ra^*_Q}^{0.4}$			&0.9553		&0.0259 \\
$Ro_{zonal}=1.05\,({Ra^*_Q}Pm^{-1/3})^{0.43}$		&0.9699		&0.0226\\
$Lo/f^{1/2}_{ohm}=0.65\,{Ra^*_Q}^{0.35}$ 		&0.9941		&0.0064 \\
$Lo/f^{1/2}_{ohm}=0.51\,{Ra^*_Q}^{0.33}Pm^{0.11}$	&0.9975		&0.0039 \\
\bottomrule
\end{tabular}
\end{table}

\section{\label{sec:conclu}Discussion and conclusions}
In this article  we investigated the effect of free-slip mechanical boundaries on various scaling laws in spherical shell dynamos. We compared the inferred scaling laws with earlier reported scalings for rigid boundary dynamos. 

We observed bistability, i.e. dipolar and multipolar dynamos coexisting for same control parameters. This agrees with the earlier findings \citep{Simitev2009, Simitev2012, Sasaki2011, Schrinner2012, Gastine2012dyn} and reinforces the importance of free-slip boundaries and zonal flows for this phenomenon. Our solar system giant planets are expected to have low local Rossby numbers ($Ro_l < 0.1$)~\citep{Olson2006}. Noting that free-slip boundaries are more appropriate for modeling these giant planets, bistability could be the reason why Jupiter and Saturn have dipole dominated magnetic fields while Uranus and Neptune have multipolar magnetic fields.

The modified Nusselt number scales as $Nu^*=0.061\,{Ra^*_Q}^{0.52}$  which is very close to the scaling for rigid boundary dynamos \citep{Christensen2006} and non-magnetic convection in spherical shell with free-slip boundaries \citep{Christensen2002}. At values of the Rayleigh number that are higher than in our simulations, a gradual transition to a weaker dependence of the Nusselt number on the Rayleigh number in expected \citep{King2009, King2010, Schmitz2010}, which can be associated with a change from a rotationally-dominated regime to a non-rotating regime. A matter of debate is whether the relative thickness of Ekman-layer and  thermal boundary layer plays a role in this transition \citep{King2009}. \cite{Schmitz2010} dispute this boundary layer hypothesis because they observe that the transition occurs similarly for both rigid and free-slip boundaries. Because our simulations do not reach the transition point, they can not contribute to this ongoing discussion.

The Rossby number scales as $Ro=0.73\,{Ra^*_Q}^{0.39}$  for dipolar and as $Ro=1.79\,{Ra^*_Q}^{0.44}$ for multipolar dynamos. The offset in the scaling of dipolar and multipolar dynamos can be attributed to different zonal flow characteristics: zonal flow quenching is stronger in a dipolar magnetic field configuration as compared to a multipolar configuration. This results in a smaller pre-factor in the Rossby number scaling for dipolar dynamos. We also used earlier numerical results of hydrodynamic convection in spherical shells with free-slip boundaries \citep{Christensen2002} and observed that the non-zonal flow scaling ($Ro_{non-zonal}$) is unaffected by the presence of magnetic field while the zonal flow scaling ($Ro_{zonal}$) is effectively quenched by magnetic field.

The corrected Lorentz number scales as $Lo/f^{1/2}_{ohm}=1.08\,{Ra^*_Q}^{0.37}$ for dipolar dynamos and $Lo/f^{1/2}_{ohm}=0.65\,{Ra^*_Q}^{0.35}$  for multipolar dynamos. The exponents are almost identical but the pre-factors differ. We investigated the origin of such shifted scaling and found that, using the scalings for Rossby number (inferred from our data-set) and an inverse relationship between ohmic dissipation time and magnetic Reynolds number, parallel and shifted scalings for $Lo/f^{1/2}_{ohm}$ are indeed expected. The observed and the expected ratio of dipolar and multipolar $Lo/f^{1/2}_{ohm}$ scalings agreed quite well. This agreement suggests that zonal flow amplitude controls the final mean magnetic field strength of the dynamos with free-slip boundaries.

Similar to \cite{Christensen2004} and \cite{Christensen2006}, we also observed that a small dependence on the magnetic Prandtl number improves the quality of the scalings, especially in the dipolar dynamo cases. However, results form the Karlsruhe laboratory dynamo experiment  motivated \cite{Christensen2004} to conjecture that such small $Pm$ dependence might disappear when $Pm<<1$. The lowest $Pm$ in our study is of order unity, which makes it difficult to ascertain this conjecture. So far, our free-slip simulations and the rigid boundary simulation results of \cite{Christensen2006} support scalings which have some $Pm$ contribution. Simulations which attain $Pm<<1$ will shed more light on this issue.

Objects such as stars and giant planets have free-surface flows,   very high density stratification, and probably fully convective interiors.  Nonetheless, scaling laws inferred from Boussinesq dynamo models with rigid boundaries that have been tailored to model the geodynamo, have been applied with some success also to giant planets and rapidly rotating stars~\citep{Christensen2009, Christensen2010}. Similar scaling of physical properties despite such drastic physical differences is puzzling. As a first step toward testing the scaling laws for conditions that are more applicable to giant planets and stars, we studied here the influence of the mechanical boundary conditions. Our analysis shows that the boundary conditions do not substantially affect the scaling behavior of the rms velocity and the magnetic field strength, which supports the validity of the original scaling laws for a broader class of objects. Future simulations of dynamos with density stratification and fully convective interiors will address the remaining critical factors.

\section*{Acknowledgements}
We are grateful to Martin Schrinner for providing extra information about his free-slip dynamo simulations. We thank Julien Aubert and another anonymous referee for very stimulating and helpful comments. We acknowledge funding from the Deutsche Forschungsgemeinschaft (DFG) through Project SFB 943 / A17 and through the special priority program 1488 (PlanetMag, http://www.planetmag.de) of the DFG. All the figures were generated using {\it matplotlib} (www.matplotlib.org).


\appendix
\section{Simulation data}

\begin{table*}
\centering
\caption{\label{tab:table2}Results for Prandtl number $Pr=1$ dynamo simulations. The $Pm$ values of bistable multipolar dynamos are marked with ``*". The critical Rayleigh numbers $Ra_c$ at which fluid convection is first excited are: $4.99\times 10^{4}$ ($E=1\times 10^{-3}$), $1.86\times 10^{5}$ ($E=3\times 10^{-4}$), $6.51\times 10^{5}$ ($E=1\times 10^{-4}$), $2.68\times 10^{6}$ ($E=3\times 10^{-5}$), $1.03\times 10^{7}$ ($E=1\times 10^{-5}$).}
\begin{tabular}{cllccccccc}
\toprule
$E$	&$Pm$	&$Ra/Ra_c$	&$Nu$	&$Rm$	&$Ro_l$	&$Ro_{zonal}$	&$Lo$	&$f_{dip}$	&$f_{ohm}$\\
\midrule
$10^{-3}$ &5 &10.02 &2.42 &322.14 &0.116 &0.0477 &0.0210 &0.0078 &0.15 \\
 &5 &20.04 &4.12 &515.39 &0.253 &0.0540 &0.0454 &0.0050 &0.21 \\
 &5 &30.06 &5.21 &701.42 &0.353 &0.0540 &0.0603 &0.0030 &0.24 \\
 &5 &40.08 &6.68 &1223.29 &0.375 &0.0966 &0.0744 &0.0120 &0.26 \\
 &5 &50.10 &7.23 &1412.75 &0.411 &0.1253 &0.0819 &0.0055 &0.25 \\
 &2 &40.08 &7.03 &635.74 &0.388 &0.1866 &0.0190 &0.0070 &0.03 \\
 &2 &50.10 &7.57 &685.70 &0.442 &0.1971 &0.0329 &0.0064 &0.07 \\
\midrule
$3\times 10^{-4}$ &10 &6.99 &2.18 &376.64 &0.038 &0.0033 &0.0302 &0.4193 &0.61 \\
 &5 &8.06 &2.27 &230.73 &0.048 &0.0041 &0.0253 &0.6320 &0.57 \\
 &5* &8.06 &2.14 &297.35 &0.057 &0.0099 &0.0115 &0.0151 &0.25 \\
 &5 &10.75 &2.86 &324.21 &0.080 &0.0041 &0.0265 &0.4800 &0.48 \\
 &5 &13.44 &3.41 &489.15 &0.108 &0.0138 &0.0203 &0.0036 &0.31 \\
 &5 &16.13 &3.94 &562.90 &0.128 &0.0151 &0.0238 &0.0050 &0.34 \\
 &3 &8.06 &2.20 &142.04 &0.051 &0.0037 &0.0212 &0.7620 &0.50 \\
 &3* &8.06 &2.09 &185.08 &0.053 &0.0117 &0.0103 &0.0121 &0.23 \\
 &3 &11.29 &2.86 &270.97 &0.080 &0.0167 &0.0156 &0.0130 &0.28 \\
 &3 &14.52 &3.57 &335.20 &0.108 &0.0206 &0.0197 &0.0127 &0.31 \\
 &1.5 &8.60 &2.02 &110.82 &0.046 &0.0177 &0.0086 &0.0180 &0.19 \\
 &1.5 &9.14 &2.20 &116.79 &0.053 &0.0179 &0.0101 &0.0240 &0.21 \\
 &1.5 &9.68 &2.37 &121.74 &0.060 &0.0180 &0.0114 &0.0210 &0.23 \\
 &1.5 &10.22 &2.52 &129.47 &0.065 &0.0189 &0.0123 &0.0270 &0.24 \\
 &1.5 &10.75 &2.60 &138.66 &0.067 &0.0206 &0.0128 &0.0120 &0.25 \\
 &1 &10.75 &2.32 &103.27 &0.054 &0.0265 &0.0099 &0.0068 &0.20 \\
 &1 &13.44 &3.12 &119.57 &0.082 &0.0277 &0.0155 &0.0173 &0.27 \\
 &0.5 &18.82 &3.99 &89.35 &0.106 &0.0451 &0.0178 &0.0188 &0.24 \\
\midrule
$10^{-4}$ &1 &7.68 &2.09 &62.39 &0.034 &0.0013 &0.0079 &0.8741 &0.42 \\
 &1 &8.45 &2.26 &69.76 &0.038 &0.0015 &0.0087 &0.8329 &0.43 \\
 &1 &9.22 &2.42 &75.82 &0.042 &0.0017 &0.0096 &0.8004 &0.43 \\
 &1* &9.22 &2.28 &96.80 &0.036 &0.0062 &0.0058 &0.0086 &0.29 \\
 &1 &12.29 &2.99 &98.48 &0.058 &0.0023 &0.0123 &0.7786 &0.46 \\
 &1* &12.29 &3.11 &130.97 &0.057 &0.0076 &0.0079 &0.0180 &0.32 \\
 &0.5 &15.36 &3.60 &64.81 &0.078 &0.0028 &0.0126 &0.8891 &0.41 \\
 &0.5* &15.36 &3.74 &84.98 &0.066 &0.0111 &0.0092 &0.0115 &0.34 \\
 &0.5 &30.72 &5.91 &109.98 &0.131 &0.0056 &0.0210 &0.7598 &0.46 \\
 &0.5 &61.44 &10.28 &203.61 &0.234 &0.0137 &0.0294 &0.5634 &0.43 \\
\midrule
$3\times 10^{-5}$ &1 &9.33 &2.41 &104.84 &0.025 &0.0007 &0.0046 &0.6392 &0.49 \\
 &0.5 &9.33 &2.37 &53.38 &0.027 &0.0006 &0.0039 &0.9071 &0.39 \\
 &0.5 &11.19 &2.81 &65.25 &0.035 &0.0008 &0.0047 &0.8317 &0.40 \\
 &0.5* &11.19 &2.78 &80.30 &0.032 &0.0026 &0.0030 &0.0001 &0.32 \\
\midrule
$10^{-5}$ &0.2 &14.56 &3.44 &49.68 &0.032 &0.0006 &0.0023 &0.9088 &0.31 \\
\bottomrule
\end{tabular}
\end{table*}


\end{document}